\documentclass[showpacs,preprintnumbers,amsmath,amssymb,prl,twocolumn]{revtex4}

\usepackage{placeins}
\usepackage{graphicx}
\usepackage{bm}
\usepackage{dcolumn}
\usepackage{array}
\usepackage{color}

\newcommand{\triplet}{a \, ^3\Sigma^+}
\newcommand{\singlet}{X \, ^1\Sigma^+}

\newcommand{\singletex}{A \, ^1\Sigma^+ - b \,^3\Pi_{0^+}}
\newcommand{\bpione}{b \,^3\Pi_{1}}

\begin{document}

\title{Ultracold dense samples of dipolar RbCs molecules \\ in the
rovibrational and hyperfine ground state}
\author{Tetsu Takekoshi$^{1,2}$}
\author{Lukas Reichs\"{o}llner$^{1}$}
\author{Andreas Schindewolf$^{1}$}
\author{Jeremy M. Hutson$^{3}$}
\author{C.~Ruth~Le~Sueur$^{3}$}
\author{Olivier Dulieu$^{4}$}
\author{Francesca Ferlaino$^{1}$}
\author{Rudolf Grimm$^{1,2}$}
\author{Hanns-Christoph N\"{a}gerl$^{1}$}
\affiliation{
$^{1}$Institut f\"ur Experimentalphysik, Universit\"at Innsbruck,
6020 Innsbruck, Austria
\\
$^{2}$Institut f\"ur Quantenoptik und Quanteninformation,
\"Osterreichische Akademie der Wissenschaften, 6020 Innsbruck, Austria
\\
$^{3}$Joint Quantum Centre (JQC) Durham/Newcastle, Department of Chemistry,
Durham University, South Road, Durham, DH1 3LE, United Kingdom
\\
$^{4}$Laboratoire Aim\'e Cotton, CNRS, Universit\'e Paris-Sud, Bat. 505,
91405 Orsay Cedex, France
}

\date{\today}

\begin{abstract}
We produce ultracold dense trapped samples of $^{87}$Rb$^{133}$Cs molecules in
their rovibrational ground state, with full nuclear hyperfine state control, by
stimulated Raman adiabatic passage (STIRAP) with efficiencies of $90\%$. We
observe the onset of hyperfine-changing collisions when the magnetic field is
ramped so that the molecules are no longer in the hyperfine ground state. A
strong quadratic shift of the transition frequencies as a function of applied
electric field shows the strongly dipolar character of the RbCs ground-state
molecule. Our results open up the prospect of realizing stable bosonic dipolar
quantum gases with ultracold molecules.
\end{abstract}

\pacs{05.30.Rt, 33.20.-t, 42.62.Fi, 67.85.-d \hfill DOI: 10.1103/PhysRevLett.113.205301}
\maketitle

Samples of ultracold molecules with dipole moments that can be tuned with
applied electric fields offer a platform for exploring many new areas of
physics. They are good candidates to form many-body systems with features such
as supersolidity, unconventional forms of superfluidity, and novel types of
quantum magnetism \cite{Trefzger2011udg, Baranov2012cmt, Lahaye2009tpo}. They
allow exquisite control over all quantum degrees of freedom, and offer the
possibility of implementing quantum simulation protocols \cite{Bloch2012qsw}
that require genuine long-range interactions.


The most advanced experiments with ultracold polar molecules to date have
been on KRb. Ni {\em et al.}\ \cite{Ni2008ahp} produced ultracold
$^{40}$K$^{87}$Rb molecules in states very close to dissociation by tuning a
magnetic field across a Feshbach resonance, and transferred the resulting
Feshbach molecules to the rovibrational absolute ground state by stimulated
Raman adiabatic passage (STIRAP). Similar work has been carried out on
non-dipolar Cs$_2$ \cite{Danzl2008qgo, Danzl2010auh}. The ground-state KRb
molecules can be transferred between hyperfine states using microwave radiation
\cite{Ospelkaus2010cth} and confined in one-dimensional \cite{deMiranda:2011}
and three-dimensional \cite{Chotia:2012} optical lattices. However, pairs of
KRb molecules can undergo an exothermic chemical reaction to form K$_2$ +
Rb$_2$; this provides an opportunity for studies of quantum state-controlled
reactions \cite{Ospelkaus2010cth, Ni2010dco,deMiranda:2011}, but also
constitutes a loss mechanism for the trapped molecules.

There is great interest in producing samples of ultracold dipolar molecules
that are collisionally stable. \.Zuchowski and Hutson \cite{Zuchowski2010rou}
have shown that the molecules NaK, NaRb, NaCs, KCs and RbCs in their absolute
ground states are stable to all possible 2-body collision processes. We have
previously demonstrated that $^{87}$Rb$^{133}$Cs Feshbach molecules can be
produced from ultracold atoms by magneto-association \cite{Debatin2011msf,
Takekoshi2012gad}. Similar work has been reported by K\"oppinger {\em et al.}\
\cite{Koeppinger2014poo}. In this paper we describe the transfer of these
molecules to their rovibrational ground state by STIRAP. We also demonstrate
magnetic control and show that the resulting molecules decay much more slowly
when they are in their hyperfine ground state than when they are in an excited
hyperfine state.

\begin{figure}[tbp]
\includegraphics[width=\columnwidth]{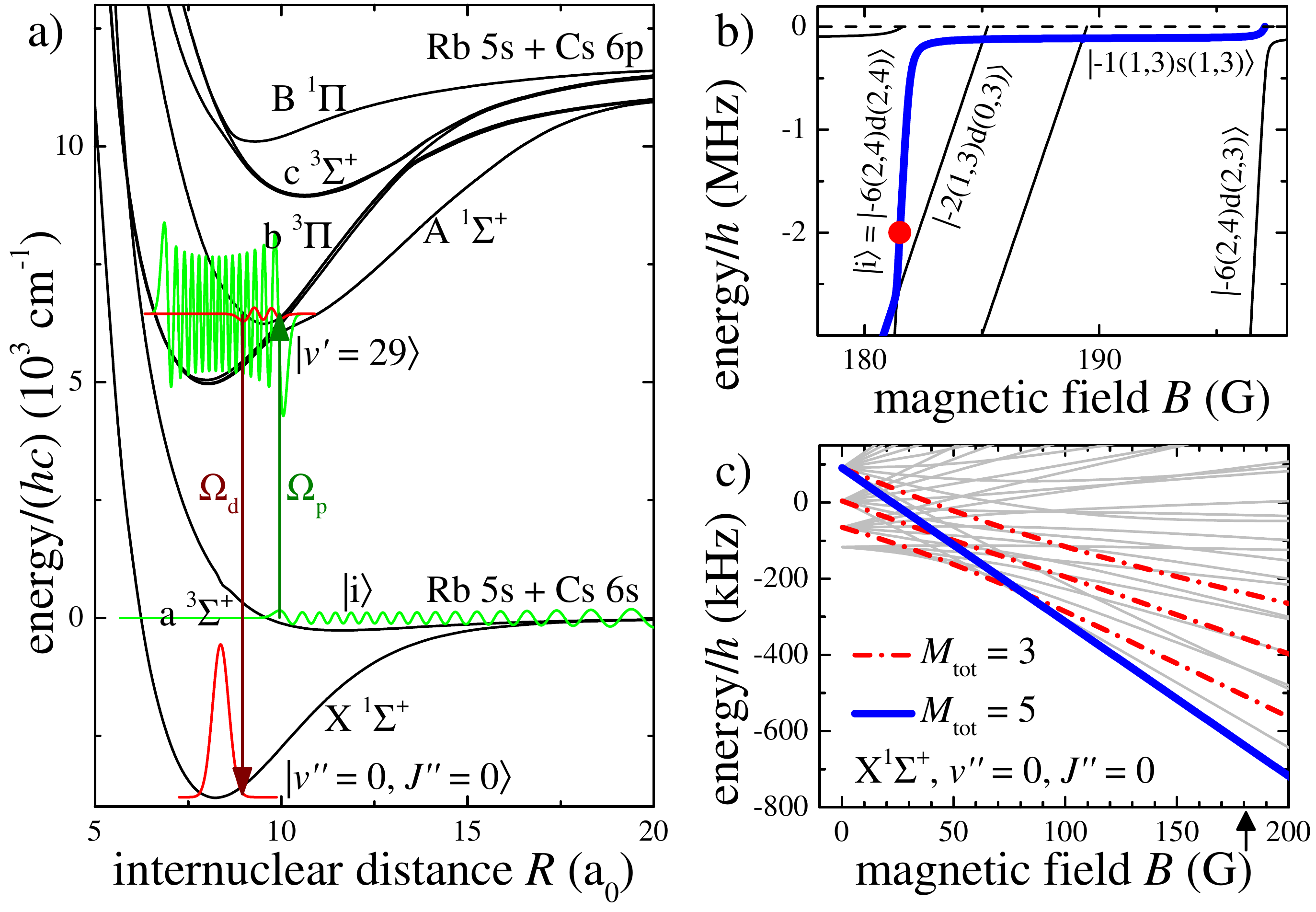}
\caption{(Color online). STIRAP scheme and levels involved. a) Ground- and
excited-state molecular potentials of the RbCs molecule \cite{Debatin2011msf}.
The transfer from the Feshbach state $|i\rangle$ at threshold to the
rovibrational ground-state level $|v''\!=\!0,J''\!=\!0\rangle$ involves the
$v'\!=\!29$ level belonging to the $\bpione$ electronically
excited state. The red and green solid lines indicate the wavefunctions that
are coupled by the STIRAP pump and dump lasers L$_\text{p}$ and L$_\text{d}$
with Rabi frequencies $\Omega_\text{p}$ and $\Omega_\text{d}$. b) Zeeman
diagram for the states with $M_{\rm tot}=4$ just below the ground-state
two-atom $(f_{\rm Rb},f_{\rm Cs})\!=\!(1,3)$ threshold. The red dot marks the
position from which STIRAP takes place. The magneto-association path is marked
with a blue line. Energies are given relative to the field-dependent atomic
dissociation threshold. c) Zeeman diagram showing the ground-state hyperfine
structure (32 states). The magnetic field during STIRAP is indicated by the
arrow. The energy levels are calculated using the Hamiltonian and parameters
from Ref.~\cite{Aldegunde2008hel}. The thick lines show the final states
allowed by the selection rule $\Delta M_{\rm tot}\!=\!\pm 1$ for vertical pump
and horizontal dump polarization (v$_\text{p}$, h$_\text{d}$).}
\label{Fig1}
\end{figure}

The states and transitions involved in our ground-state molecule production
process are shown in Fig.~\ref{Fig1} (a)-(c). A pump laser beam L$_\text{p}$
at $1557$~nm couples a Feshbach state $|i\rangle$ with mostly $\triplet$
character to the $|v'\!=\!29\rangle$ level of the $\bpione$ state with Rabi
frequency $\Omega_\text{p}$. This state has a small admixture of the
$A^1\Sigma^+$ state \cite{Debatin2011msf,supmat}, and a dump laser beam
L$_\text{d}$ at $977$~nm couples it to the rovibrational ground-state level
$|v''\!=\!0,J''\!=\!0\rangle$ of the $\singlet$ potential with Rabi frequency
$\Omega_\text{d}$. This level is made up of 32 Zeeman sublevels, as shown in
Fig.~\ref{Fig1}(c) \cite{Aldegunde2009hel}. At $B=0$ the levels are grouped
according to the total molecular nuclear spin $I''=2$, 3, 4, or 5. The
stretched state with $M_{I''} \!=\! M_{\rm tot}\!=\!5$ is
the absolute ground state for $B$ larger than about $90$~G. It can be accessed
at $B=181$~G using crossed vertical and horizontal linear polarizations
(v$_\text{p}$, h$_\text{d}$) for L$_\text{p}$ and L$_\text{d}$ co-propagating
in the horizontal plane.

We start by generating a sample of $^{87}$Rb$^{133}$Cs Feshbach molecules via
magneto-association in an ultracold, magnetically levitated and nearly
quantum-degenerate mixture of Rb and Cs atoms. The molecules are initially
produced using the Feshbach resonance at $B\!=\!197.06$~G and then transferred
by magnetic field ramps to the state $|{-2}(1,3)d(0,3)\rangle$ near
$B\!=\!180$~G as sketched in Fig.~\ref{Fig1}(b) and described in more detail
in Ref.~\cite{Takekoshi2012gad}. Here, states are labeled with quantum numbers
$|n(f_{\rm Rb},f_{\rm Cs})L(m_{f_{\rm Rb}},m_{f_{\rm Cs}})\rangle$, where $n$
is the vibrational quantum number counted downwards from the
$(f_{\rm Rb},f_{\rm Cs})$ dissociation threshold, $f$ indicates the atomic
total angular momentum with projection $m_f$, and $L$ is the molecular
rotational angular momentum. We take the quantization axis to lie along the
magnetic field direction, which is vertical in our setup \cite{supmat}.

The high-field-seeking molecules in state $|{-2}(1,3)d(0,3)\rangle$ are
separated from the remaining atoms by the Stern-Gerlach effect. The magnetic
field $B$ is then ramped back up through the nearest avoided crossing to
transfer the molecules into the strongly low-field-seeking state
$|i\rangle\!=\!|{-6}(2,4)d(2,4)\rangle$ at a binding energy of approximately
2~MHz$\times h$ at $B\!=\!181$~G (marked with a dot in Fig.~\ref{Fig1}(b)).
This state is chosen because it has the greatest triplet fraction and the
largest amplitude at short range, giving the most favorable Franck-Condon
overlap for the STIRAP process described below. To reduce spatial Zeeman
broadening and gravitational sag, the field gradient used for levitation is
turned off and a vertical 1D optical lattice \cite{supmat} is superimposed on
the molecular cloud to hold it against gravity. The molecular sample is thus
held in a stack of pancake-shaped 2D traps with their tight axis along the
vertical direction. This additional step, combined with the
shorter collisional lifetime of molecules in the $n=-6$ state (about $30$~ms),
reduces the cloud population from 3000 to between 1000 and 1500 trapped
molecules with a $1/e^2$-cloud radius of between 30 and 40~$\mu$m. The
translational temperature measured in expansion after sudden release from the
trap is 240(30)~nK. The overall sample preparation procedure takes about
$13$~s.

\begin{figure}[tbp]
\includegraphics[width=\columnwidth]{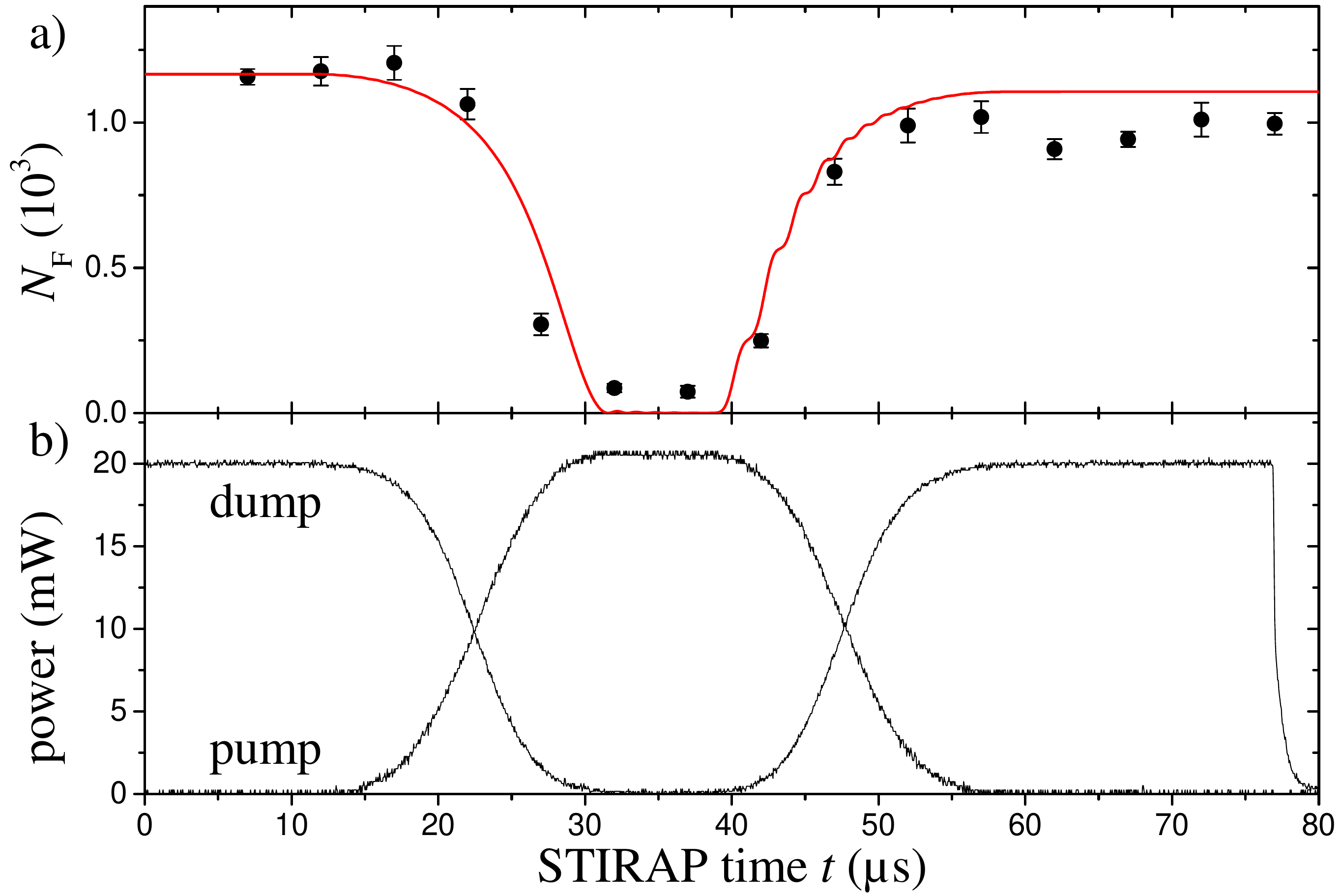}
\caption{(Color online). Efficient ground-state STIRAP transfer. a) Number of
Feshbach molecules $N_\text{F}$ as a function of STIRAP time $t$ during a
typical forward and reverse on-resonance STIRAP pulse sequence as shown in b).
The peak Rabi frequencies are $\Omega_\text{p} \!=\! 2\pi \times 0.77 (22)$ MHz
and $\Omega_\text{d} \!=\! 2\pi \times 2.3(6)$ MHz. The one-way STIRAP
efficiency is 90\%. The red curve is the result of a master equation model
\cite{supmat}. Error bars denote the $1\sigma$ standard statistical error.
b) Laser power as a function of time $t$ as recorded by photodiodes.}
\label{Fig2}
\end{figure}

\begin{figure}[tbp]
\includegraphics[width=\columnwidth]{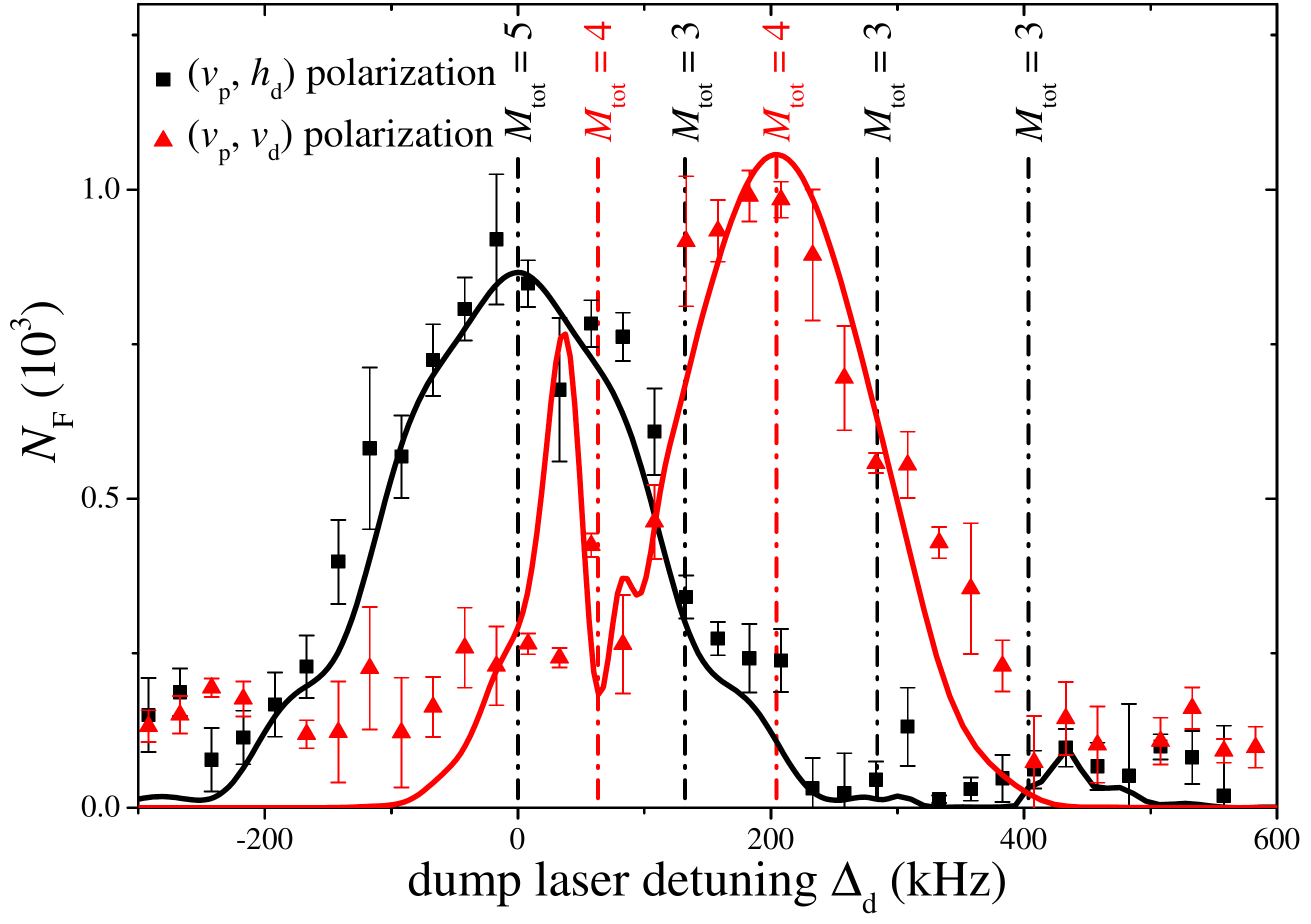}
\caption{(Color online). STIRAP spectrum showing the number of Feshbach
molecules $N_\text{F}$ after round-trip STIRAP as a function of dump laser
detuning $\Delta_\text{d}$ for two different choices of the polarization of
the dump laser. The energies of the hyperfine components of the ground state
(calculated using the parameters of Ref.~\cite{Aldegunde2008hel}) are marked
with dashed vertical lines and labeled with their total angular momentum
projection $M_{\rm tot}$. For (v$_\text{p}$, h$_\text{d}$) polarization (black
squares) the Feshbach molecules ($M_{\rm tot}\!=\!4$) are primarily transferred
into the absolute hyperfine ground state with $M_{\rm tot}\!=\!5$. For
(v$_\text{p}$, v$_\text{d}$) polarization (red triangles), hyperfine-excited
levels are addressed. The solid curves are master equation simulation results
\cite{supmat}. The black curve centered around zero detuning is a fit to the
data to determine the dump Rabi frequency. The Rabi frequencies are
$\Omega_\text{p} \!=\! 2\pi \times 0.26(7)$ MHz and $\Omega_\text{d} \!=\! 2\pi
\times 2.3(6)$ MHz.}
\label{Fig3}
\end{figure}

STIRAP is based on a pulse sequence in which the dump laser is turned on
before the pump laser to generate a transient dark superposition of the
initial and final states \cite{Bergmann1998cpt}. We perform ground-state
STIRAP from $|i\rangle$ to $|v''\!=\!0,J''\!=\!0\rangle$ and characterize
its efficiency by reversing the STIRAP process as shown in Fig.~\ref{Fig2}
\cite{Debatin2013PhD}. Molecules are transferred to the hyperfine-Zeeman
ground state with $M_{\rm tot}\!=\!5$ between $t\!\approx\!15$ and $30 \
\mu$s and back to the Feshbach state $|i\rangle$ between $t\!\approx\!40$ and
55~$\mu$s. Both lasers are tuned to one-photon resonance for fixed
$B\!=\!181$~G. The Feshbach molecules are then detected by dissociating them
at the Feshbach resonance at $197.06$~G and using absorption imaging on the
atomic clouds \cite{Takekoshi2012gad}. The round-trip transfer efficiencies
are typically about $80\%$, implying one-way transfer efficiencies of about
$90\%$. For comparison, the solid line in Fig.~\ref{Fig2}(a) is the result of
a simulation that takes laser linewidth into account, but not beam shape and
laser noise pedestal effects \cite{supmat}. It gives a somewhat higher
efficiency.

Scanning the dump laser detuning $\Delta_\text{d}$ reveals
hyperfine and Zeeman substructure of the $\singlet$,
$|v''\!=\!0,J''\!=\!0\rangle$ state as shown in Fig.~\ref{Fig3}. For
(v$_\text{p}$, h$_\text{d}$) polarization the transfer is mostly into the
level with $M_{\rm tot}\!=\!5$. For (v$_\text{p}$, v$_\text{d}$) polarization
the transfer is primarily into one of the two hyperfine-excited levels with
$M_{\rm tot}\!=\!4$. The most important terms in the ground-state hyperfine
Hamiltonian \cite{Aldegunde2008hel} are the nuclear Zeeman shift and the scalar
nuclear spin-spin interaction, which are governed by the electronic and nuclear
$g$ factors, and the nuclear spin-spin parameter $c_4$, respectively. The
second of these two terms dominates at low field. The $g$ factors are very
precisely known, and a simulation \cite{supmat} using them and the $c_4$
parameter of Ref.~\cite{Aldegunde2008hel} agrees well with the observed
spectrum for both choices of polarisation.

There are usually 50 to 100 Feshbach molecules that remain after the transfer
to the ground state (offset in Figs.~\ref{Fig2}(a) and \ref{Fig3}). We
believe this is mainly due to a slight beam misalignment and the fact that the
molecular cloud and STIRAP beams have similar radii. We exclude these molecules
when calculating the transfer efficiency. The efficiency is most likely limited
by laser power, in the sense that Feshbach molecules at the edge of the cloud
see lower laser intensities. Laser phase noise pedestals may also play a role,
as discussed in Ref.~\cite{PhysRevA.89.013831}.

\begin{figure}[tbp]
\includegraphics[width=\columnwidth]{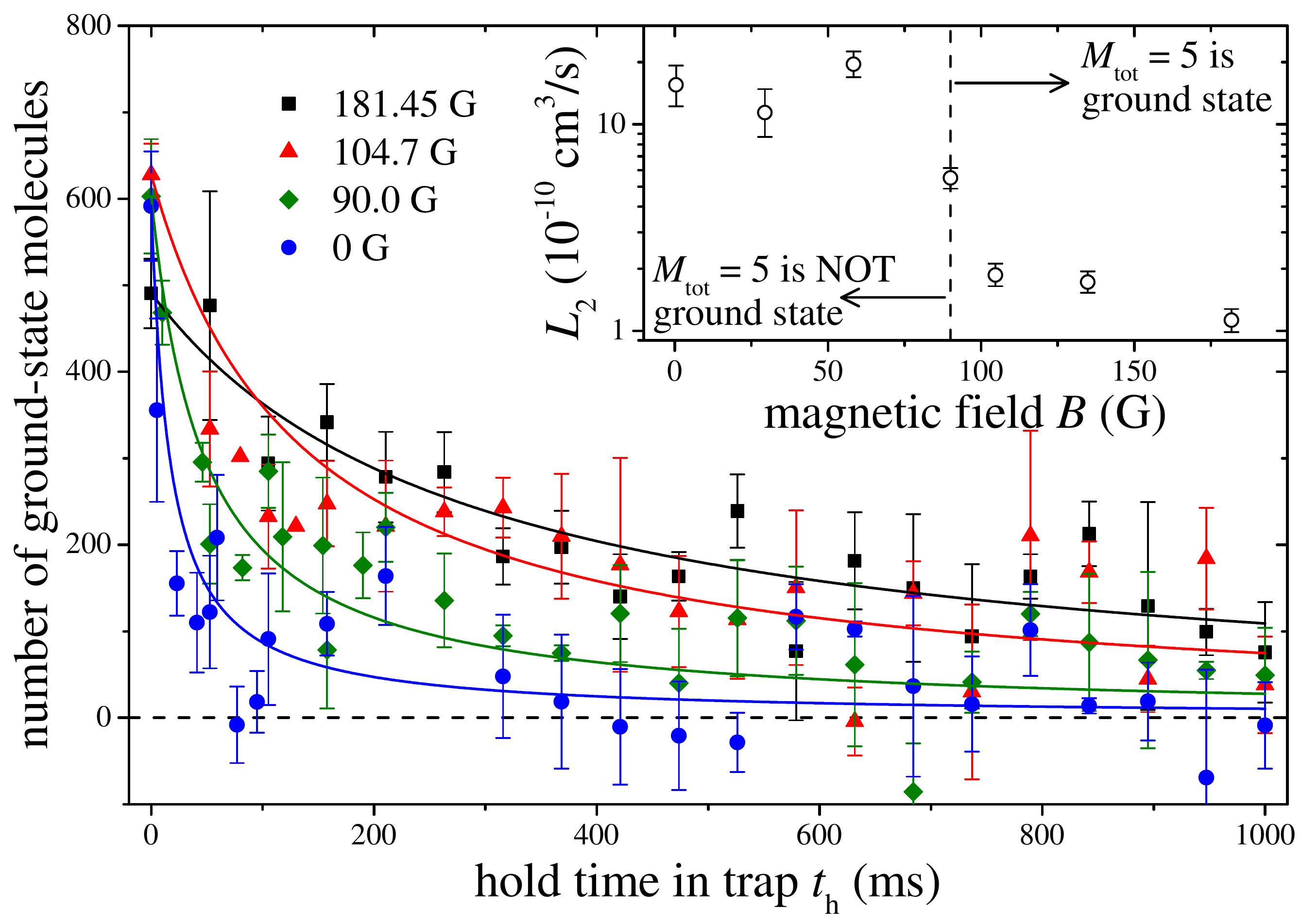}
\caption{(Color online). Decay of ground-state molecules as a result of
collisions at zero electric field. The number of ground-state molecules in $M_{\rm tot}\!=\!5$ is
plotted against hold time $t_\text{h}$ in the crossed dipole trap for
different values of the magnetic field $B$ as indicated. The initial peak
density is $1.1(1)\times10^{11}$ cm$^{-3}$. The solid lines are fits based on
a two-body decay model to determine the two-body loss rate coefficient $L_2$
\cite{supmat}. The fits are constrained to run through the first data point at
zero hold time. The inset plots $L_2$ as a function of $B$. A greatly reduced
$L_2$ is seen at magnetic fields $B$ greater than about 90~G, where the
molecules are mostly in the hyperfine-Zeeman ground state.}
\label{Fig4}
\end{figure}

To explore the molecules' collisional properties we load our sample of
ground-state molecules into a three-dimensional crossed dipole trap
\cite{supmat}. The trap is comparatively stiff with a geometrically averaged
trap frequency of $ 409(20) $ Hz to hold the sample against gravity. The
sample's peak particle density is now $1.1(1)\times10^{11}$ cm$^{-3}$. The
compression of the sample leads to a marked increase in temperature to
$8.7(7) \ \mu$K. Nevertheless, we expect that $s-$wave collisions still
dominate the collision process. Fig.~\ref{Fig4} shows the ground-state
population in the $M_{\rm tot}\!=\!5$ state as a function of hold time
$t_\text{h}$ between forward and reverse STIRAP transfer for various values of
the magnetic field $B$. For this measurement, we first prepare the molecular
sample as before at $B=181$~G in $M_{\rm tot}\!=\!5$ and then ramp the magnetic
field to the chosen value within about $1$~ms. After time $t_\text{h}$ we
reverse the process and determine the remaining number of molecules. The
results show ground-state molecule loss that depends strongly on $B$. Using a
two-body decay model \cite{supmat} we determine the two-body loss rate
coefficient $L_2$. Its dependence on $B$ is shown in the inset to
Fig.~\ref{Fig4}. The value of $L_2$ is considerably greater at fields below
about $90$~G. The state with $M_{\rm tot}\!=\!5$ is not the absolute ground
state at fields below this threshold, as seen in Fig.~\ref{Fig1}(c), and we
attribute the greatly reduced lifetime to hyperfine-changing collisions to form
the lower-energy states. We note that $L_2$ is non-zero even for fields above
90 G; this may be due to thermal population of excited hyperfine states, or to
losses involving long-lived collision complexes
\cite{Mayle2013sou,Cs2tobepublished}. We also note that our ground-state sample
is not 100\% pure, because it initially contains some molecules left behind in
the Feshbach state $|i\rangle$. The cross section for inelastic collisions
between molecules in states $|M_{\rm tot}\!=\!5\rangle$ and $|i\rangle$ is
likely to be large, and will lead to some loss of ground-state molecules on
the timescales considered here.



\begin{figure}[tbp]
\includegraphics[width=\columnwidth]{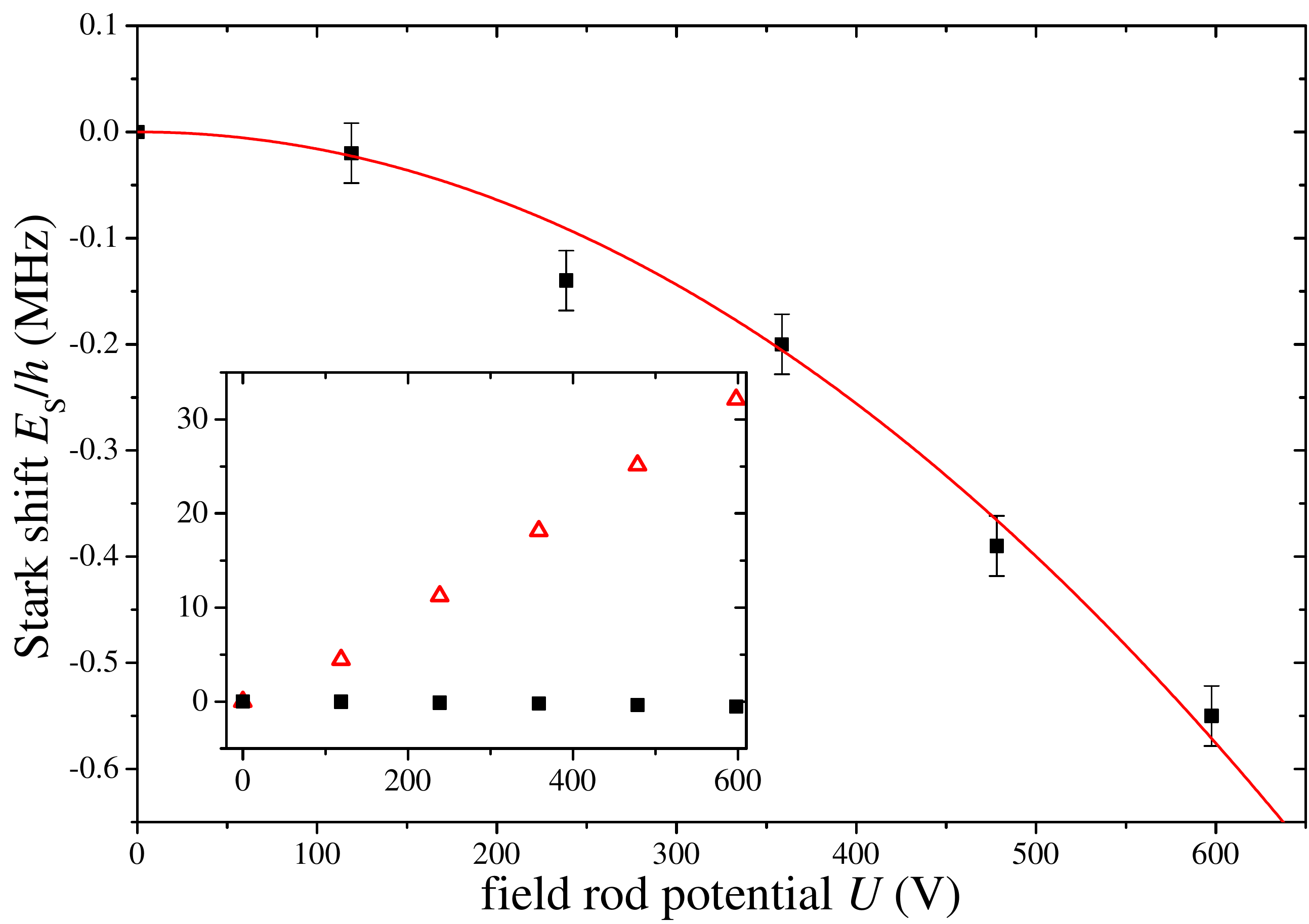}
\caption{(Color online). Stark shift of the $M_{\rm tot}\!=\!5$ state of
the RbCs ground state. The two-photon STIRAP resonance shift $E_\text{S}$ is
plotted as a function of the electrode potential $U$. The solid line is a
quadratic fit. The inset shows an expanded range in which the excited-state
shift can be seen as well (triangles).}
\label{Fig5}
\end{figure}

A crucial property of RbCs molecules is their permanent electric dipole moment
$\mu$, calculated to be $1.25$~D  in the absolute ground state
\cite{Aymar2005coa,Kotochigova2005air}. We have measured the Stark shift of
the hyperfine ground state by applying voltages to a set of four parallel
electrodes external to the fused silica cell vacuum chamber \cite{supmat} and
tracking the shift $E_\text{S}$ of the $M_{\rm tot}\!=\!5$ peak position (as
in Fig.~\ref{Fig3}) from that recorded at zero electrode potential. The
potential is pulsed to reduce charging effects from the alkali-coated cell
walls \cite{supmat,Bouchiat1999eco}. The resulting shift is shown in
Fig.~\ref{Fig5}. Both the dump and the pump laser must be detuned
considerably, because of the large excited-state shift shown in the inset of
Fig.~\ref{Fig5}. The quadratic shift is observed to be $1.60(7)$
$\rm{Hz/V^{2}}$, which implies a permanent dipole moment of $1.17(2)(4)$~D.
Here, the first error is statistical, the second is the estimated systematic
error due to geometrical uncertainty that enters when calculating the
dielectrically enhanced electric field inside our quartz cell apparatus
\cite{supmat}.

In conclusion, we have formed dense samples of ultracold RbCs molecules in
their electronic and rovibrational ground state. The molecules are initially
formed in near-dissociation states by magneto-association, and transferred to
the ground state by the STIRAP method. The efficiency of the ground-state
transfer is about 90\%. With an appropriate choice of laser polarization, we
can produce the molecules in their absolute hyperfine ground state. RbCs
molecules in their ground state are stable to all possible two-body collision
processes, so our results offer the prospect of producing the first
collisionally stable quantum gas of dipolar molecules.

In future work, we will attempt to increase the sample size and density by
creating Feshbach molecules from atomic Mott insulators in a
three-dimensional optical lattice \cite{Lercher2011poa}, in generalization of
work on homonuclear Cs$_2$ \cite{Danzl2010auh}. The dynamics will then be
dominated by nearest-neighbor interactions with interaction strength on the
order of $h\!\times\!1$~kHz. This will allow us to study important problems
in quantum many-body physics, such as the phase diagram of the Bose-Hubbard
model extended by a long-range interaction term \cite{Damski2003coa,
Capogrosso2010qpo}.

We acknowledge contributions by V. Pramhaas, M. Kugler, and M. Debatin and
thank N. Bouloufa, R. Vexiau, A. Crubellier and J. Aldegunde for fruitful
discussions. We acknowledge support by the Austrian Science Fund (FWF) through
the Spezialforschungsbereich (SFB) FoQuS within project P06 (FWF project
number F4006-N23), the European Office of Aerospace Research and Development
through Grant FA8655-10-1-3033 and the Engineering and Physical Sciences
Research Council through Grant no.\ EP/I012044/1.


\newpage
\clearpage

\section{Supplemental Material}

\subsection{Molecular states involved}
In contrast to the case of homonuclear alkali dimers, whose excited states have
a resonant dipole-dipole van der Waals interaction, in heteronuclear dimers the
excited and ground states both have $1/R^6$ van der Waals interactions, where
$R$ is the internuclear distance. This makes it more likely that a suitable
excited state exists that simultaneously has good Franck-Condon overlap with
the Feshbach and ground states, making four-photon transfer schemes unnecessary
\cite{Danzl2010auh2}.

We initially chose laser wavelengths to address states of the strongly coupled
$\singletex$ system but found the ground-state transfer efficiency to be
extremely poor, even though the predicted and measured laser couplings were
strong. A transition to the $\bpione$, $|v'\!=\!29\rangle$ state
\cite{Debatin2011msf2} worked much better and the results using this
intermediate state are presented in this work. We attribute the low
$\singletex$ transfer efficiency to the fact that Zeeman and hyperfine
splittings in $\Omega'\!=\!0$ states are much smaller than our one-photon
STIRAP linewidth ($\sim \Omega_\text{p0} \Omega_\text{d0} \tau_\text{p}$, where
$\Omega_\text{p0}$ and $\Omega_\text{d0}$ are the peak pump and dump Rabi
frequencies and $\tau_\text{p}$ is the STIRAP-pulse overlap time
\cite{Vitanov1997pos}). In this case, multiple intermediate states are
addressed simultaneously, which causes loss \cite{Shore1995cpt, Martin1995cpt,
Vitanov1999apt}. While this can often be remedied by detuning far from all
intermediate states, large Rabi frequencies are often required. In STIRAP of
Cs$_2$ \cite{Danzl2010auh2}, where the magnetic field is low enough that the
total angular momentum $F$ is nearly a good quantum number, a fixed ratio
between the dump and pump Rabi frequencies of the multiple intermediate states
exists and the losses are ameliorated \cite{Vitanov1999apt}.

Decomposing the Feshbach state $|i\rangle$ into its Hund's case (a)
nuclear-spin decoupled basis states reveals it to be a superposition of
$J\!=\!1$, 2 and 3 with only the $J\!=\!1$ component coupling to
$\bpione , J'\!=\!1$ due to the selection rules $\Delta J\!=\!0,\pm1$ and
$\Delta \Sigma\!=\!0$. Here $J$ is the total molecular angular momentum
excluding nuclear spin and $\Sigma$ is the projection of the total electron
spin onto the internuclear axis. The intermediate $\bpione$ state is actually
not expected to allow coupling to the $\singlet$ ground state due to the
selection rule $\Delta S\!=\!0$. However, the $v'\!=\!29$, $J'\!=\!1$ level of
$\bpione$ system lies only 16~GHz away \cite{Debatin2011msf2} from the
$v'\!=\!38$, $J'\!=\!1$ level of the the $\singletex$ system and contains a
small admixture of $A \, ^1\Sigma^+$, $J'=1$ \cite{Docenko2010sds}, which can
be coupled to the $\singlet, J''\!=\!0$ ground state. The calculated
vibrational wavefunctions for both intermediate components are indicated in
Fig.~1 of the main text. A small admixture of $B^1\Pi$ is also possible and
cannot be ruled out \cite{Svetlana}. The $\bpione, J'\!=\!1$ hyperfine and
Zeeman manifold spans about $h\times700$~MHz and we empirically found the state
with the strongest couplings using $\pi$ (vertically) polarized pump light and
$\sigma^+ \!-\! \sigma^-$ (horizontally) polarized dump light.

At $B\!=\!181$~G the absolute ground state is the stretched $M_{\rm tot}\!=\!5$
state. This has the same nuclear spin orientation $m_{i_{\rm Rb}}\!=\!3/2$,
$m_{i_{\rm Cs}}\!=\!7/2$ as the initial Feshbach state, implying that the
intermediate state must also have high $m_{i_{\rm Rb}}\!=\!3/2$,
$m_{i_{\rm Cs}}\!=\!7/2$ character for good transfer efficiency. This is indeed
the case. Subsequent spectroscopy and analysis of the $\bpione$, $v'=29$,
$J'=1$ state \cite{tobepublished} has revealed that the lowest of the 18
allowed transitions of the intermediate state manifold gives the strongest
one-photon coupling and has quantum numbers $M_{\rm tot}\!=\!4$, $J'\!=\!1$,
$m_{J'}\!\approx\!-1$, $m_{i_{\rm Rb}}\!\approx\!3/2$, and
$m_{i_{\rm Cs}}\!\approx\!7/2$. Our intermediate state model gives us a
complete picture of the possibilities for direct transfer to other
$\singlet , J''\!=\!0$ states should the need arise.

\subsection{Laser light generation}
The pump and dump STIRAP light is derived from diode lasers that are stabilized
via phase locks using additional transfer diode lasers and tunable
radiofrequency sources to two independent high-finesse cavities
\cite{Alnis2008sld} with finesse $F=240,000$ and $440,000$, respectively. Each
laser can be tuned by hundreds of MHz with radiofrequency precision and
delivers a little over $20$~mW into a beam with a $1/e^2$ intensity radius of
about $40$~$\mu$m at the position of the molecular sample. The short-term laser
linewidths inferred from the lock signals are approximately 90 and 170~Hz and
the measured broadening due to acoustic effects in the optical fibers used for
beam delivery is less than $70$~Hz.

\subsection{Optical lattice}
The 1D optical lattice is generated by narrow-band laser light at
$\lambda = 1064.5$~nm propagating along the vertical $z$-direction. The light
with a power up to $P=0.8$~W is collimated to a beam with $1/e^2$-radius of
$415\ \mu$m at the position of the molecular sample. The depth of the lattice
is about $V_{z}= 48 E_\text{R}^\mathrm{RbCs}$ at $P=0.34$~W, where
$E_\text{R}^\mathrm{RbCs}=h^2/(2m_\mathrm{RbCs} \lambda^2) = h \times 0.8 $ kHz
is the photon recoil energy, with $m_\mathrm{RbCs}$ the mass of the RbCs
molecule. At such a depth, tunneling from well to well is fully suppressed.

\subsection{Modeling the STIRAP time course}

To simulate the STIRAP process we use a master equation in Lindblad form
\cite{GardinerZoller2010}. The effects of finite laser linewidths are included
in the model, but the effects of beam shape and phase noise pedestals are not.
The basis set contains the Feshbach state, the excited state, and the lowest 10
hyperfine and Zeeman ground states. All parameters except the Franck-Condon
factor for the dump transition are either known or have been measured. This
factor is varied until the black theory curve fits the data near the main
(v$_\text{p}$, h$_\text{d}$) peak shown in Fig.~3 of the main text.

From the natural linewidth of the excited level $|v'\!=\!29\rangle$, measured
to be $135(10)$~kHz, along with a one-photon on-resonance absorption
measurement \cite{Takekoshi2012gad2}, we obtain a pump Rabi frequency of
$\Omega_\text{p} \!=\! 2\pi \times
0.84(24)$kHz$\sqrt{I_\text{p}/(\rm{mW}/\rm{cm}^2)}$, where $I_\text{p}$ is the
pump laser intensity. We estimate the dump Rabi frequency by varying it until
the ground-state transfer simulation roughly agrees with the (v$_\text{p}$,
h$_\text{d}$) polarization data (Fig.~3). This gives a dump Rabi frequency of
$\Omega_\text{d} \!=\! 2\pi \times
2.76(67)$kHz$\sqrt{I_\text{d}/(\rm{mW}/\rm{cm}^2)}$, where $I_\text{d}$ is the
pump laser intensity.

\subsection{Crossed dipole trap and two-body decay model}
We perform the collisional measurements in a crossed dipole trap whose trap
frequencies are $(\omega_x,\omega_y,\omega_z)= 2\pi \times (439,294,529) $ Hz
in the horizontal $(x,y)$ and vertical $(z)$ directions. For modeling the decay
we assume a Gaussian spatial distribution for a sample that remains in (quasi)
thermal equilibrium at temperature $T$. The rate equation for the density of
ground-state molecules $n_\text{g}$ reads
\begin{equation*}
  \dot{n}_\text{g} = - L_\text{2} n_\text{g}^2
\end{equation*}
with the two-body loss rate coefficient $L_\text{2}$. Here, we assume that the
decay is dominated by a two-body process. Introducing the effective volume
$V_\text{eff} = (m\bar{\omega}^2/(4\pi k_\text{B}T))^{-3/2}$, where
$\bar{\omega}$ is the geometrically averaged trap frequency $\bar{\omega}
= (\omega_\text{x}\omega_\text{y}\omega_\text{z})^{1/2}$, allows us to rewrite
this equation in the form
\begin{equation*}
  \dot{N}_\text{g} = - (L_\text{2} / V_\text{eff}) N_\text{g}^2,
\end{equation*}
where $N_\text{g}$ is the number of ground-state molecules. We fit solutions to
this equation to the data shown in Fig.~4 to determine $L_2$ as shown in the
inset to Fig.~4. We note that an attempt to model our data assuming just a
one-body loss process fails a chi-squared test.

\subsection{DC-Stark shift measurement setup}
We apply an electric field by putting voltages on four nearly parallel 8.0
mm-diameter 140 mm-long external rod electrodes at the corners of the
rectangular-cross-sectioned fused silica vacuum cell. The rods are separated by
$52.0(1)$~mm along the vertical direction and by $68(1)$~mm horizontally. The
position of the molecular sample with respect to the electrodes is know to
about $1$ mm. We estimate the field from this geometry to be $60(3)\%$ of the
value expected from infinitely wide parallel plates separated by $52.0$~mm.


\begin{thebibliography}{29}
\expandafter\ifx\csname natexlab\endcsname\relax\def\natexlab#1{#1}\fi
\expandafter\ifx\csname bibnamefont\endcsname\relax
  \def\bibnamefont#1{#1}\fi
\expandafter\ifx\csname bibfnamefont\endcsname\relax
  \def\bibfnamefont#1{#1}\fi
\expandafter\ifx\csname citenamefont\endcsname\relax
  \def\citenamefont#1{#1}\fi
\expandafter\ifx\csname url\endcsname\relax
  \def\url#1{\texttt{#1}}\fi
\expandafter\ifx\csname urlprefix\endcsname\relax\def\urlprefix{URL }\fi
\providecommand{\bibinfo}[2]{#2}
\providecommand{\eprint}[2][]{\url{#2}}

\bibitem[{\citenamefont{Trefzger et~al.}(2011)\citenamefont{Trefzger, Menotti,
  Capogrosso-Sansone, and Lewenstein}}]{Trefzger2011udg}
\bibinfo{author}{\bibfnamefont{C.}~\bibnamefont{Trefzger}},
  \bibinfo{author}{\bibfnamefont{C.}~\bibnamefont{Menotti}},
  \bibinfo{author}{\bibfnamefont{B.}~\bibnamefont{Capogrosso-Sansone}},
  \bibnamefont{and}
  \bibinfo{author}{\bibfnamefont{M.}~\bibnamefont{Lewenstein}},
  \bibinfo{journal}{J. Phys. B: At. Mol. Opt. Phys.}
  \textbf{\bibinfo{volume}{44}}, \bibinfo{pages}{193001}
  (\bibinfo{year}{2011}).

\bibitem[{\citenamefont{Baranov et~al.}(2012)\citenamefont{Baranov, Dalmonte,
  Pupillo, and Zoller}}]{Baranov2012cmt}
\bibinfo{author}{\bibfnamefont{M.}~\bibnamefont{Baranov}},
  \bibinfo{author}{\bibfnamefont{M.}~\bibnamefont{Dalmonte}},
  \bibinfo{author}{\bibfnamefont{G.}~\bibnamefont{Pupillo}}, \bibnamefont{and}
  \bibinfo{author}{\bibfnamefont{P.}~\bibnamefont{Zoller}},
  \bibinfo{journal}{Chem. Rev.} \textbf{\bibinfo{volume}{112}},
  \bibinfo{pages}{5012} (\bibinfo{year}{2012}).

\bibitem[{\citenamefont{Lahaye et~al.}(2009)\citenamefont{Lahaye, Menotti,
  Santos, Lewenstein, and Pfau}}]{Lahaye2009tpo}
\bibinfo{author}{\bibfnamefont{T.}~\bibnamefont{Lahaye}},
  \bibinfo{author}{\bibfnamefont{C.}~\bibnamefont{Menotti}},
  \bibinfo{author}{\bibfnamefont{L.}~\bibnamefont{Santos}},
  \bibinfo{author}{\bibfnamefont{M.}~\bibnamefont{Lewenstein}},
  \bibnamefont{and} \bibinfo{author}{\bibfnamefont{T.}~\bibnamefont{Pfau}},
  \bibinfo{journal}{Rep. Prog. Phys.} \textbf{\bibinfo{volume}{72}},
  \bibinfo{pages}{126401} (\bibinfo{year}{2009}).

\bibitem[{\citenamefont{Bloch et~al.}(2012)\citenamefont{Bloch, Dalibard, and
  Nascimb\`ene}}]{Bloch2012qsw}
\bibinfo{author}{\bibfnamefont{I.}~\bibnamefont{Bloch}},
  \bibinfo{author}{\bibfnamefont{J.}~\bibnamefont{Dalibard}}, \bibnamefont{and}
  \bibinfo{author}{\bibfnamefont{S.}~\bibnamefont{Nascimb\`ene}},
  \bibinfo{journal}{Nature Phys.} \textbf{\bibinfo{volume}{8}},
  \bibinfo{pages}{267} (\bibinfo{year}{2012}).

\bibitem[{\citenamefont{Ni et~al.}(2008)\citenamefont{Ni, Ospelkaus,
  de~Miranda, Pe'er, Neyenhuis, Zirbel, Kotochigova, Julienne, Jin, and
  Ye}}]{Ni2008ahp}
\bibinfo{author}{\bibfnamefont{K.-K.} \bibnamefont{Ni}},
  \bibinfo{author}{\bibfnamefont{S.}~\bibnamefont{Ospelkaus}},
  \bibinfo{author}{\bibfnamefont{M.~H.~G.} \bibnamefont{de~Miranda}},
  \bibinfo{author}{\bibfnamefont{A.}~\bibnamefont{Pe'er}},
  \bibinfo{author}{\bibfnamefont{B.}~\bibnamefont{Neyenhuis}},
  \bibinfo{author}{\bibfnamefont{J.~J.} \bibnamefont{Zirbel}},
  \bibinfo{author}{\bibfnamefont{S.}~\bibnamefont{Kotochigova}},
  \bibinfo{author}{\bibfnamefont{P.~S.} \bibnamefont{Julienne}},
  \bibinfo{author}{\bibfnamefont{D.~S.} \bibnamefont{Jin}}, \bibnamefont{and}
  \bibinfo{author}{\bibfnamefont{J.}~\bibnamefont{Ye}},
  \bibinfo{journal}{Science} \textbf{\bibinfo{volume}{322}},
  \bibinfo{pages}{231} (\bibinfo{year}{2008}).

\bibitem[{\citenamefont{Danzl et~al.}(2008)\citenamefont{Danzl, Haller,
  Gustavsson, Mark, Hart, Bouloufa, Dulieu, Ritsch, and
  N\"agerl}}]{Danzl2008qgo}
\bibinfo{author}{\bibfnamefont{J.~G.} \bibnamefont{Danzl}},
  \bibinfo{author}{\bibfnamefont{E.}~\bibnamefont{Haller}},
  \bibinfo{author}{\bibfnamefont{M.}~\bibnamefont{Gustavsson}},
  \bibinfo{author}{\bibfnamefont{M.~J.} \bibnamefont{Mark}},
  \bibinfo{author}{\bibfnamefont{R.}~\bibnamefont{Hart}},
  \bibinfo{author}{\bibfnamefont{N.}~\bibnamefont{Bouloufa}},
  \bibinfo{author}{\bibfnamefont{O.}~\bibnamefont{Dulieu}},
  \bibinfo{author}{\bibfnamefont{H.}~\bibnamefont{Ritsch}}, \bibnamefont{and}
  \bibinfo{author}{\bibfnamefont{H.-C.} \bibnamefont{N\"agerl}},
  \bibinfo{journal}{Science} \textbf{\bibinfo{volume}{321}},
  \bibinfo{pages}{1062} (\bibinfo{year}{2008}).

\bibitem[{\citenamefont{Danzl et~al.}(2010)\citenamefont{Danzl, Mark, Haller,
  Gustavsson, Hart, Aldegunde, Hutson, and N\"{a}gerl}}]{Danzl2010auh}
\bibinfo{author}{\bibfnamefont{J.~G.} \bibnamefont{Danzl}},
  \bibinfo{author}{\bibfnamefont{M.~J.} \bibnamefont{Mark}},
  \bibinfo{author}{\bibfnamefont{E.}~\bibnamefont{Haller}},
  \bibinfo{author}{\bibfnamefont{M.}~\bibnamefont{Gustavsson}},
  \bibinfo{author}{\bibfnamefont{R.}~\bibnamefont{Hart}},
  \bibinfo{author}{\bibfnamefont{J.}~\bibnamefont{Aldegunde}},
  \bibinfo{author}{\bibfnamefont{J.~M.} \bibnamefont{Hutson}},
  \bibnamefont{and} \bibinfo{author}{\bibfnamefont{H.-C.}
  \bibnamefont{N\"{a}gerl}}, \bibinfo{journal}{Nature Phys.}
  \textbf{\bibinfo{volume}{6}}, \bibinfo{pages}{265} (\bibinfo{year}{2010}).

\bibitem[{\citenamefont{Ospelkaus et~al.}(2010)\citenamefont{Ospelkaus, Ni,
  Qu\'em\'ener, Neyenhuis, Wang, de~Miranda, Bohn, Ye, and
  Jin}}]{Ospelkaus2010cth}
\bibinfo{author}{\bibfnamefont{S.}~\bibnamefont{Ospelkaus}},
  \bibinfo{author}{\bibfnamefont{K.-K.} \bibnamefont{Ni}},
  \bibinfo{author}{\bibfnamefont{G.}~\bibnamefont{Qu\'em\'ener}},
  \bibinfo{author}{\bibfnamefont{B.}~\bibnamefont{Neyenhuis}},
  \bibinfo{author}{\bibfnamefont{D.}~\bibnamefont{Wang}},
  \bibinfo{author}{\bibfnamefont{M.~H.~G.} \bibnamefont{de~Miranda}},
  \bibinfo{author}{\bibfnamefont{J.~L.} \bibnamefont{Bohn}},
  \bibinfo{author}{\bibfnamefont{J.}~\bibnamefont{Ye}}, \bibnamefont{and}
  \bibinfo{author}{\bibfnamefont{D.~S.} \bibnamefont{Jin}},
  \bibinfo{journal}{Phys. Rev. Lett.} \textbf{\bibinfo{volume}{104}},
  \bibinfo{pages}{030402} (\bibinfo{year}{2010}).

\bibitem[{\citenamefont{de~Miranda et~al.}(2011)\citenamefont{de~Miranda,
  Chotia, Neyenhuis, Wang, Qu\'em\'ener, Ospelkaus, Bohn, Ye, and
  Jin}}]{deMiranda:2011}
\bibinfo{author}{\bibfnamefont{M.~H.~G.} \bibnamefont{de~Miranda}},
  \bibinfo{author}{\bibfnamefont{A.}~\bibnamefont{Chotia}},
  \bibinfo{author}{\bibfnamefont{B.}~\bibnamefont{Neyenhuis}},
  \bibinfo{author}{\bibfnamefont{D.}~\bibnamefont{Wang}},
  \bibinfo{author}{\bibfnamefont{G.}~\bibnamefont{Qu\'em\'ener}},
  \bibinfo{author}{\bibfnamefont{S.}~\bibnamefont{Ospelkaus}},
  \bibinfo{author}{\bibfnamefont{J.~L.} \bibnamefont{Bohn}},
  \bibinfo{author}{\bibfnamefont{J.}~\bibnamefont{Ye}}, \bibnamefont{and}
  \bibinfo{author}{\bibfnamefont{D.~S.} \bibnamefont{Jin}},
  \bibinfo{journal}{Nature Phys.} \textbf{\bibinfo{volume}{7}},
  \bibinfo{pages}{502} (\bibinfo{year}{2011}).

\bibitem[{\citenamefont{Chotia et~al.}(2012)\citenamefont{Chotia, Neyenhuis,
  Moses, Yan, Covey, Foss-Feig, Rey, Jin, and Ye}}]{Chotia:2012}
\bibinfo{author}{\bibfnamefont{A.}~\bibnamefont{Chotia}},
  \bibinfo{author}{\bibfnamefont{B.}~\bibnamefont{Neyenhuis}},
  \bibinfo{author}{\bibfnamefont{S.~A.} \bibnamefont{Moses}},
  \bibinfo{author}{\bibfnamefont{B.}~\bibnamefont{Yan}},
  \bibinfo{author}{\bibfnamefont{J.~P.} \bibnamefont{Covey}},
  \bibinfo{author}{\bibfnamefont{M.}~\bibnamefont{Foss-Feig}},
  \bibinfo{author}{\bibfnamefont{A.~M.} \bibnamefont{Rey}},
  \bibinfo{author}{\bibfnamefont{D.~S.} \bibnamefont{Jin}}, \bibnamefont{and}
  \bibinfo{author}{\bibfnamefont{J.}~\bibnamefont{Ye}}, \bibinfo{journal}{Phys.
  Rev. Lett.} \textbf{\bibinfo{volume}{108}}, \bibinfo{pages}{080405}
  (\bibinfo{year}{2012}).

\bibitem[{\citenamefont{Ni et~al.}(2010)\citenamefont{Ni, Ospelkaus, Wang,
  Quemener, Neyenhuis, {de Miranda}, Bohn, Ye, and Jin}}]{Ni2010dco}
\bibinfo{author}{\bibfnamefont{K.-K.} \bibnamefont{Ni}},
  \bibinfo{author}{\bibfnamefont{S.}~\bibnamefont{Ospelkaus}},
  \bibinfo{author}{\bibfnamefont{D.}~\bibnamefont{Wang}},
  \bibinfo{author}{\bibfnamefont{G.}~\bibnamefont{Quemener}},
  \bibinfo{author}{\bibfnamefont{B.}~\bibnamefont{Neyenhuis}},
  \bibinfo{author}{\bibfnamefont{M.~H.~G.} \bibnamefont{{de Miranda}}},
  \bibinfo{author}{\bibfnamefont{J.~L.} \bibnamefont{Bohn}},
  \bibinfo{author}{\bibfnamefont{J.}~\bibnamefont{Ye}}, \bibnamefont{and}
  \bibinfo{author}{\bibfnamefont{D.~S.} \bibnamefont{Jin}},
  \bibinfo{journal}{Nature} \textbf{\bibinfo{volume}{464}},
  \bibinfo{pages}{1324} (\bibinfo{year}{2010}).

\bibitem[{\citenamefont{{\.{Z}}uchowski and Hutson}(2010)}]{Zuchowski2010rou}
\bibinfo{author}{\bibfnamefont{P.~S.} \bibnamefont{{\.{Z}}uchowski}}
  \bibnamefont{and} \bibinfo{author}{\bibfnamefont{J.~M.}
  \bibnamefont{Hutson}}, \bibinfo{journal}{Phys. Rev. A}
  \textbf{\bibinfo{volume}{81}}, \bibinfo{pages}{060703(R)}
  (\bibinfo{year}{2010}).

\bibitem[{\citenamefont{Debatin et~al.}(2011)\citenamefont{Debatin, Takekoshi,
  Rameshan, Reich\-s\"ollner, Ferlaino, Grimm, Vexiau, Bouloufa, Dulieu, and
  N\"agerl}}]{Debatin2011msf}
\bibinfo{author}{\bibfnamefont{M.}~\bibnamefont{Debatin}},
  \bibinfo{author}{\bibfnamefont{T.}~\bibnamefont{Takekoshi}},
  \bibinfo{author}{\bibfnamefont{R.}~\bibnamefont{Rameshan}},
  \bibinfo{author}{\bibfnamefont{L.}~\bibnamefont{Reich\-s\"ollner}},
  \bibinfo{author}{\bibfnamefont{F.}~\bibnamefont{Ferlaino}},
  \bibinfo{author}{\bibfnamefont{R.}~\bibnamefont{Grimm}},
  \bibinfo{author}{\bibfnamefont{R.}~\bibnamefont{Vexiau}},
  \bibinfo{author}{\bibfnamefont{N.}~\bibnamefont{Bouloufa}},
  \bibinfo{author}{\bibfnamefont{O.}~\bibnamefont{Dulieu}}, \bibnamefont{and}
  \bibinfo{author}{\bibfnamefont{H.-C.} \bibnamefont{N\"agerl}},
  \bibinfo{journal}{Phys. Chem. Chem. Phys.} \textbf{\bibinfo{volume}{13}},
  \bibinfo{pages}{18926} (\bibinfo{year}{2011}).

\bibitem[{\citenamefont{Takekoshi et~al.}(2012)\citenamefont{Takekoshi,
  Debatin, Rameshan, Ferlaino, Grimm, N\"agerl, Le~Sueur, Hutson, Julienne,
  Kotochigova et~al.}}]{Takekoshi2012gad}
\bibinfo{author}{\bibfnamefont{T.}~\bibnamefont{Takekoshi}},
  \bibinfo{author}{\bibfnamefont{M.}~\bibnamefont{Debatin}},
  \bibinfo{author}{\bibfnamefont{R.}~\bibnamefont{Rameshan}},
  \bibinfo{author}{\bibfnamefont{F.}~\bibnamefont{Ferlaino}},
  \bibinfo{author}{\bibfnamefont{R.}~\bibnamefont{Grimm}},
  \bibinfo{author}{\bibfnamefont{H.-C.} \bibnamefont{N\"agerl}},
  \bibinfo{author}{\bibfnamefont{C.~R.} \bibnamefont{Le~Sueur}},
  \bibinfo{author}{\bibfnamefont{J.~M.} \bibnamefont{Hutson}},
  \bibinfo{author}{\bibfnamefont{P.~S.} \bibnamefont{Julienne}},
  \bibinfo{author}{\bibfnamefont{S.}~\bibnamefont{Kotochigova}},
  \bibnamefont{et~al.}, \bibinfo{journal}{Phys. Rev. A}
  \textbf{\bibinfo{volume}{85}}, \bibinfo{pages}{032506}
  (\bibinfo{year}{2012}).

\bibitem[{\citenamefont{K\"oppinger et~al.}(2014)\citenamefont{K\"oppinger,
  McCarron, Jenkin, Molony, Cho, Cornish, Le~Sueur, Blackley, and
  Hutson}}]{Koeppinger2014poo}
\bibinfo{author}{\bibfnamefont{M.~P.} \bibnamefont{K\"oppinger}},
  \bibinfo{author}{\bibfnamefont{D.~J.} \bibnamefont{McCarron}},
  \bibinfo{author}{\bibfnamefont{D.~L.} \bibnamefont{Jenkin}},
  \bibinfo{author}{\bibfnamefont{P.~K.} \bibnamefont{Molony}},
  \bibinfo{author}{\bibfnamefont{H.-W.} \bibnamefont{Cho}},
  \bibinfo{author}{\bibfnamefont{S.~L.} \bibnamefont{Cornish}},
  \bibinfo{author}{\bibfnamefont{C.~R.} \bibnamefont{Le~Sueur}},
  \bibinfo{author}{\bibfnamefont{C.~L.} \bibnamefont{Blackley}},
  \bibnamefont{and} \bibinfo{author}{\bibfnamefont{J.~M.}
  \bibnamefont{Hutson}}, \bibinfo{journal}{Phys. Rev. A}
  \textbf{\bibinfo{volume}{89}}, \bibinfo{pages}{033604}
  (\bibinfo{year}{2014}).

\bibitem[{sup()}]{supmat}
\bibinfo{note}{See Supplemental Material
  for details on the molecular states, on laser light generation, on the
  optical lattice, on modeling the STIRAP time course, on the two-body decay
	model, and on the DC-Stark shift measurement setup.}

\bibitem[{\citenamefont{Aldegunde and Hutson}(2009)}]{Aldegunde2009hel}
\bibinfo{author}{\bibfnamefont{J.}~\bibnamefont{Aldegunde}} \bibnamefont{and}
  \bibinfo{author}{\bibfnamefont{J.~M.} \bibnamefont{Hutson}},
  \bibinfo{journal}{Phys. Rev. A} \textbf{\bibinfo{volume}{79}},
  \bibinfo{pages}{013401} (\bibinfo{year}{2009}).

\bibitem[{\citenamefont{Aldegunde et~al.}(2008)\citenamefont{Aldegunde,
  Rivington, \.{Z}uchowski, and Hutson}}]{Aldegunde2008hel}
\bibinfo{author}{\bibfnamefont{J.}~\bibnamefont{Aldegunde}},
  \bibinfo{author}{\bibfnamefont{B.~A.} \bibnamefont{Rivington}},
  \bibinfo{author}{\bibfnamefont{P.~S.} \bibnamefont{\.{Z}uchowski}},
  \bibnamefont{and} \bibinfo{author}{\bibfnamefont{J.~M.}
  \bibnamefont{Hutson}}, \bibinfo{journal}{Phys. Rev. A}
  \textbf{\bibinfo{volume}{78}}, \bibinfo{pages}{033434}
  (\bibinfo{year}{2008}).

\bibitem[{\citenamefont{Bergmann et~al.}(1998)\citenamefont{Bergmann, Theuer,
  and Shore}}]{Bergmann1998cpt}
\bibinfo{author}{\bibfnamefont{K.}~\bibnamefont{Bergmann}},
  \bibinfo{author}{\bibfnamefont{H.}~\bibnamefont{Theuer}}, \bibnamefont{and}
  \bibinfo{author}{\bibfnamefont{B.~W.} \bibnamefont{Shore}},
  \bibinfo{journal}{Rev. Mod. Phys.} \textbf{\bibinfo{volume}{70}},
  \bibinfo{pages}{1003} (\bibinfo{year}{1998}).

\bibitem[{\citenamefont{Debatin}(2013)}]{Debatin2013PhD}
\bibinfo{author}{\bibfnamefont{M.}~\bibnamefont{Debatin}}, Ph.D. thesis,
  \bibinfo{school}{University of Innsbruck} (\bibinfo{year}{2013}).

\bibitem[{\citenamefont{Yatsenko et~al.}(2014)\citenamefont{Yatsenko, Shore,
  and Bergmann}}]{PhysRevA.89.013831}
\bibinfo{author}{\bibfnamefont{L.~P.} \bibnamefont{Yatsenko}},
  \bibinfo{author}{\bibfnamefont{B.~W.} \bibnamefont{Shore}}, \bibnamefont{and}
  \bibinfo{author}{\bibfnamefont{K.}~\bibnamefont{Bergmann}},
  \bibinfo{journal}{Phys. Rev. A} \textbf{\bibinfo{volume}{89}},
  \bibinfo{pages}{013831} (\bibinfo{year}{2014}).




\bibitem[{\citenamefont{Mayle et~al.}(2013)\citenamefont{Mayle,
  Qu\'{e}m\'{e}ner, Ruzic, and Bohn}}]{Mayle2013sou}
\bibinfo{author}{\bibfnamefont{M.}~\bibnamefont{Mayle}},
  \bibinfo{author}{\bibfnamefont{G.}~\bibnamefont{Qu\'{e}m\'{e}ner}},
  \bibinfo{author}{\bibfnamefont{B.~P.} \bibnamefont{Ruzic}}, \bibnamefont{and}
  \bibinfo{author}{\bibfnamefont{J.~L.} \bibnamefont{Bohn}},
  \bibinfo{journal}{Phys. Rev. A} \textbf{\bibinfo{volume}{87}},
  \bibinfo{pages}{012709} (\bibinfo{year}{2013}).

\bibitem[{\citenamefont{Lauber et~al.}()\citenamefont{Lauber, Kiliov, Mark,
  Meinert, and N\"agerl}}]{Cs2tobepublished}
\bibinfo{author}{\bibfnamefont{K.}~\bibnamefont{Lauber}},
  \bibinfo{author}{\bibfnamefont{E.}~\bibnamefont{Kiliov}},
  \bibinfo{author}{\bibfnamefont{M.~J.} \bibnamefont{Mark}},
  \bibinfo{author}{\bibfnamefont{F.}~\bibnamefont{Meinert}}, \bibnamefont{and}
  \bibinfo{author}{\bibfnamefont{H.-C.} \bibnamefont{N\"agerl}},
  \bibinfo{note}{to be published}.

\bibitem[{\citenamefont{Aymar and Dulieu}(2005)}]{Aymar2005coa}
\bibinfo{author}{\bibfnamefont{M.}~\bibnamefont{Aymar}} \bibnamefont{and}
  \bibinfo{author}{\bibfnamefont{O.}~\bibnamefont{Dulieu}},
  \bibinfo{journal}{J. Chem. Phys.} \textbf{\bibinfo{volume}{122}},
  \bibinfo{pages}{204302} (\bibinfo{year}{2005}).

\bibitem[{\citenamefont{Kotochigova and Tiesinga}(2005)}]{Kotochigova2005air}
\bibinfo{author}{\bibfnamefont{S.}~\bibnamefont{Kotochigova}} \bibnamefont{and}
  \bibinfo{author}{\bibfnamefont{E.}~\bibnamefont{Tiesinga}},
  \bibinfo{journal}{J. Chem. Phys.} \textbf{\bibinfo{volume}{123}},
  \bibinfo{eid}{174304} (\bibinfo{year}{2005}).

\bibitem[{\citenamefont{Bouchiat et~al.}(1999)\citenamefont{Bouchiat, Guena,
  Jacquier, Lintz, and Papoyan}}]{Bouchiat1999eco}
\bibinfo{author}{\bibfnamefont{M.~A.} \bibnamefont{Bouchiat}},
  \bibinfo{author}{\bibfnamefont{J.}~\bibnamefont{Guena}},
  \bibinfo{author}{\bibfnamefont{P.}~\bibnamefont{Jacquier}},
  \bibinfo{author}{\bibfnamefont{M.}~\bibnamefont{Lintz}}, \bibnamefont{and}
  \bibinfo{author}{\bibfnamefont{A.~V.} \bibnamefont{Papoyan}},
  \bibinfo{journal}{Appl. Phys. B} \textbf{\bibinfo{volume}{68}},
  \bibinfo{pages}{1109} (\bibinfo{year}{1999}).

\bibitem[{\citenamefont{Lercher et~al.}(2011)\citenamefont{Lercher, Takekoshi,
  Debatin, Schuster, Rameshan, Ferlaino, Grimm, and N\"agerl}}]{Lercher2011poa}
\bibinfo{author}{\bibfnamefont{A.}~\bibnamefont{Lercher}},
  \bibinfo{author}{\bibfnamefont{T.}~\bibnamefont{Takekoshi}},
  \bibinfo{author}{\bibfnamefont{M.}~\bibnamefont{Debatin}},
  \bibinfo{author}{\bibfnamefont{B.}~\bibnamefont{Schuster}},
  \bibinfo{author}{\bibfnamefont{R.}~\bibnamefont{Rameshan}},
  \bibinfo{author}{\bibfnamefont{F.}~\bibnamefont{Ferlaino}},
  \bibinfo{author}{\bibfnamefont{R.}~\bibnamefont{Grimm}}, \bibnamefont{and}
  \bibinfo{author}{\bibfnamefont{H.-C.} \bibnamefont{N\"agerl}},
  \bibinfo{journal}{Eur. Phys. J. D} \textbf{\bibinfo{volume}{65}},
  \bibinfo{pages}{3} (\bibinfo{year}{2011}).

\bibitem[{\citenamefont{Damski et~al.}(2003)\citenamefont{Damski, Santos,
  Tiemann, Lewenstein, Kotochigova, Julienne, and Zoller}}]{Damski2003coa}
\bibinfo{author}{\bibfnamefont{B.}~\bibnamefont{Damski}},
  \bibinfo{author}{\bibfnamefont{L.}~\bibnamefont{Santos}},
  \bibinfo{author}{\bibfnamefont{E.}~\bibnamefont{Tiemann}},
  \bibinfo{author}{\bibfnamefont{M.}~\bibnamefont{Lewenstein}},
  \bibinfo{author}{\bibfnamefont{S.}~\bibnamefont{Kotochigova}},
  \bibinfo{author}{\bibfnamefont{P.}~\bibnamefont{Julienne}}, \bibnamefont{and}
  \bibinfo{author}{\bibfnamefont{P.}~\bibnamefont{Zoller}},
  \bibinfo{journal}{Phys. Rev. Lett.} \textbf{\bibinfo{volume}{90}},
  \bibinfo{pages}{110401} (\bibinfo{year}{2003}).

\bibitem[{\citenamefont{Capogrosso-Sansone
  et~al.}(2010)\citenamefont{Capogrosso-Sansone, Trefzger, Lewenstein, Zoller,
  and Pupillo}}]{Capogrosso2010qpo}
\bibinfo{author}{\bibfnamefont{B.}~\bibnamefont{Capogrosso-Sansone}},
  \bibinfo{author}{\bibfnamefont{C.}~\bibnamefont{Trefzger}},
  \bibinfo{author}{\bibfnamefont{M.}~\bibnamefont{Lewenstein}},
  \bibinfo{author}{\bibfnamefont{P.}~\bibnamefont{Zoller}}, \bibnamefont{and}
  \bibinfo{author}{\bibfnamefont{G.}~\bibnamefont{Pupillo}},
  \bibinfo{journal}{Phys. Rev. Lett.} \textbf{\bibinfo{volume}{104}},
  \bibinfo{pages}{125301} (\bibinfo{year}{2010}).

\end{thebibliography}

\begin{thebibliography}{12}
\expandafter\ifx\csname natexlab\endcsname\relax\def\natexlab#1{#1}\fi
\expandafter\ifx\csname bibnamefont\endcsname\relax
  \def\bibnamefont#1{#1}\fi
\expandafter\ifx\csname bibfnamefont\endcsname\relax
  \def\bibfnamefont#1{#1}\fi
\expandafter\ifx\csname citenamefont\endcsname\relax
  \def\citenamefont#1{#1}\fi
\expandafter\ifx\csname url\endcsname\relax
  \def\url#1{\texttt{#1}}\fi
\expandafter\ifx\csname urlprefix\endcsname\relax\def\urlprefix{URL }\fi
\providecommand{\bibinfo}[2]{#2}
\providecommand{\eprint}[2][]{\url{#2}}

\bibitem[{\citenamefont{Danzl et~al.}(2010)\citenamefont{Danzl, Mark, Haller,
  Gustavsson, Hart, Aldegunde, Hutson, and N\"{a}gerl}}]{Danzl2010auh2}
\bibinfo{author}{\bibfnamefont{J.~G.} \bibnamefont{Danzl}},
  \bibinfo{author}{\bibfnamefont{M.~J.} \bibnamefont{Mark}},
  \bibinfo{author}{\bibfnamefont{E.}~\bibnamefont{Haller}},
  \bibinfo{author}{\bibfnamefont{M.}~\bibnamefont{Gustavsson}},
  \bibinfo{author}{\bibfnamefont{R.}~\bibnamefont{Hart}},
  \bibinfo{author}{\bibfnamefont{J.}~\bibnamefont{Aldegunde}},
  \bibinfo{author}{\bibfnamefont{J.~M.} \bibnamefont{Hutson}},
  \bibnamefont{and} \bibinfo{author}{\bibfnamefont{H.-C.}
  \bibnamefont{N\"{a}gerl}}, \bibinfo{journal}{Nature Phys.}
  \textbf{\bibinfo{volume}{6}}, \bibinfo{pages}{265} (\bibinfo{year}{2010}).

\bibitem[{\citenamefont{Debatin et~al.}(2011)\citenamefont{Debatin, Takekoshi,
  Rameshan, Reich\-s\"ollner, Ferlaino, Grimm, Vexiau, Bouloufa, Dulieu, and
  N\"agerl}}]{Debatin2011msf2}
\bibinfo{author}{\bibfnamefont{M.}~\bibnamefont{Debatin}},
  \bibinfo{author}{\bibfnamefont{T.}~\bibnamefont{Takekoshi}},
  \bibinfo{author}{\bibfnamefont{R.}~\bibnamefont{Rameshan}},
  \bibinfo{author}{\bibfnamefont{L.}~\bibnamefont{Reich\-s\"ollner}},
  \bibinfo{author}{\bibfnamefont{F.}~\bibnamefont{Ferlaino}},
  \bibinfo{author}{\bibfnamefont{R.}~\bibnamefont{Grimm}},
  \bibinfo{author}{\bibfnamefont{R.}~\bibnamefont{Vexiau}},
  \bibinfo{author}{\bibfnamefont{N.}~\bibnamefont{Bouloufa}},
  \bibinfo{author}{\bibfnamefont{O.}~\bibnamefont{Dulieu}}, \bibnamefont{and}
  \bibinfo{author}{\bibfnamefont{H.-C.} \bibnamefont{N\"agerl}},
  \bibinfo{journal}{Phys. Chem. Chem. Phys.} \textbf{\bibinfo{volume}{13}},
  \bibinfo{pages}{18926} (\bibinfo{year}{2011}).

\bibitem[{\citenamefont{Vitanov and Stenholm}(1997)}]{Vitanov1997pos}
\bibinfo{author}{\bibfnamefont{N.~V.} \bibnamefont{Vitanov}} \bibnamefont{and}
  \bibinfo{author}{\bibfnamefont{S.}~\bibnamefont{Stenholm}},
  \bibinfo{journal}{Opt. Comm.} \textbf{\bibinfo{volume}{135}},
  \bibinfo{pages}{394} (\bibinfo{year}{1997}).

\bibitem[{\citenamefont{Shore et~al.}(1995)\citenamefont{Shore, Martin, Fewell,
  and Bergmann}}]{Shore1995cpt}
\bibinfo{author}{\bibfnamefont{B.~W.} \bibnamefont{Shore}},
  \bibinfo{author}{\bibfnamefont{J.}~\bibnamefont{Martin}},
  \bibinfo{author}{\bibfnamefont{M.~P.} \bibnamefont{Fewell}},
  \bibnamefont{and} \bibinfo{author}{\bibfnamefont{K.}~\bibnamefont{Bergmann}},
  \bibinfo{journal}{Phys. Rev. A} \textbf{\bibinfo{volume}{52}},
  \bibinfo{pages}{566} (\bibinfo{year}{1995}).

\bibitem[{\citenamefont{Martin et~al.}(1995)\citenamefont{Martin, Shore, and
  Bergmann}}]{Martin1995cpt}
\bibinfo{author}{\bibfnamefont{J.}~\bibnamefont{Martin}},
  \bibinfo{author}{\bibfnamefont{B.~W.} \bibnamefont{Shore}}, \bibnamefont{and}
  \bibinfo{author}{\bibfnamefont{K.}~\bibnamefont{Bergmann}},
  \bibinfo{journal}{Phys. Rev. A} \textbf{\bibinfo{volume}{52}},
  \bibinfo{pages}{583} (\bibinfo{year}{1995}).

\bibitem[{\citenamefont{Vitanov and Stenholm}(1999)}]{Vitanov1999apt}
\bibinfo{author}{\bibfnamefont{N.~V.} \bibnamefont{Vitanov}} \bibnamefont{and}
  \bibinfo{author}{\bibfnamefont{S.}~\bibnamefont{Stenholm}},
  \bibinfo{journal}{Phys. Rev. A} \textbf{\bibinfo{volume}{60}},
  \bibinfo{pages}{3820} (\bibinfo{year}{1999}).

\bibitem[{\citenamefont{Docenko et~al.}(2010)\citenamefont{Docenko, Tamanis, Ferber,
  Bergeman, Kotochigova, Stolyarov, de~Faria~Nogueira, and
  Fellows}}]{Docenko2010sds}
\bibinfo{author}{\bibfnamefont{O.}~\bibnamefont{Docenko}},
  \bibinfo{author}{\bibfnamefont{M.}~\bibnamefont{Tamanis}},
	\bibinfo{author}{\bibfnamefont{R.}~\bibnamefont{Ferber}},
  \bibinfo{author}{\bibfnamefont{T.}~\bibnamefont{Bergeman}},
  \bibinfo{author}{\bibfnamefont{S.}~\bibnamefont{Kotochigova}},
  \bibinfo{author}{\bibfnamefont{A.~V.} \bibnamefont{Stolyarov}},
  \bibinfo{author}{\bibfnamefont{A.}~\bibnamefont{de~Faria~Nogueira}},
  \bibnamefont{and} \bibinfo{author}{\bibfnamefont{C.~E.}
  \bibnamefont{Fellows}}, \bibinfo{journal}{Phys. Rev. A}
  \textbf{\bibinfo{volume}{81}}, \bibinfo{pages}{042511}
  (\bibinfo{year}{2010}).

\bibitem[{\citenamefont{Kotochigova}()}]{Svetlana}
\bibinfo{author}{\bibfnamefont{S.}~\bibnamefont{Kotochigova}},
  \bibinfo{note}{private communication}.

\bibitem[{\citenamefont{Takekoshi et~al.}()\citenamefont{Takekoshi, Grimm, and
  N\"agerl}}]{tobepublished}
\bibinfo{author}{\bibfnamefont{T.}~\bibnamefont{Takekoshi}},
  \bibinfo{author}{\bibfnamefont{R.}~\bibnamefont{Grimm}}, \bibnamefont{and}
  \bibinfo{author}{\bibfnamefont{H.-C.} \bibnamefont{N\"agerl}},
  \bibinfo{note}{to be published}.

\bibitem[{\citenamefont{Alnis et~al.}(2008)\citenamefont{Alnis, Matveev,
  Kolachevsky, Udem, and H\"ansch}}]{Alnis2008sld}
\bibinfo{author}{\bibfnamefont{J.}~\bibnamefont{Alnis}},
  \bibinfo{author}{\bibfnamefont{A.}~\bibnamefont{Matveev}},
  \bibinfo{author}{\bibfnamefont{N.}~\bibnamefont{Kolachevsky}},
  \bibinfo{author}{\bibfnamefont{T.}~\bibnamefont{Udem}}, \bibnamefont{and}
  \bibinfo{author}{\bibfnamefont{T.~W.} \bibnamefont{H\"ansch}},
  \bibinfo{journal}{Phys. Rev. A} \textbf{\bibinfo{volume}{77}},
  \bibinfo{pages}{053809} (\bibinfo{year}{2008}).

\bibitem[{\citenamefont{Gardiner and Zoller}(2010)}]{GardinerZoller2010}
\bibinfo{author}{\bibfnamefont{C.~W.} \bibnamefont{Gardiner}} \bibnamefont{and}
  \bibinfo{author}{\bibfnamefont{P.}~\bibnamefont{Zoller}},
  \emph{\bibinfo{title}{Quantum Noise}} (\bibinfo{publisher}{Springer,
  Berlin-Heidelberg}, \bibinfo{year}{2010}).

\bibitem[{\citenamefont{Takekoshi et~al.}(2012)\citenamefont{Takekoshi,
  Debatin, Rameshan, Ferlaino, Grimm, N\"agerl, Le~Sueur, Hutson, Julienne,
  Kotochigova et~al.}}]{Takekoshi2012gad2}
\bibinfo{author}{\bibfnamefont{T.}~\bibnamefont{Takekoshi}},
  \bibinfo{author}{\bibfnamefont{M.}~\bibnamefont{Debatin}},
  \bibinfo{author}{\bibfnamefont{R.}~\bibnamefont{Rameshan}},
  \bibinfo{author}{\bibfnamefont{F.}~\bibnamefont{Ferlaino}},
  \bibinfo{author}{\bibfnamefont{R.}~\bibnamefont{Grimm}},
  \bibinfo{author}{\bibfnamefont{H.-C.} \bibnamefont{N\"agerl}},
  \bibinfo{author}{\bibfnamefont{C.~R.} \bibnamefont{Le~Sueur}},
  \bibinfo{author}{\bibfnamefont{J.~M.} \bibnamefont{Hutson}},
  \bibinfo{author}{\bibfnamefont{P.~S.} \bibnamefont{Julienne}},
  \bibinfo{author}{\bibfnamefont{S.}~\bibnamefont{Kotochigova}},
  \bibnamefont{et~al.}, \bibinfo{journal}{Phys. Rev. A}
  \textbf{\bibinfo{volume}{85}}, \bibinfo{pages}{032506}
  (\bibinfo{year}{2012}).

\end{thebibliography}

\end{document}